\documentstyle[12pt]{article}
\begin{document}
 \bibliographystyle{unsrt}
 \vbox{\vspace{6mm}}

%\renewcommand{\baselinestretch}{2}

%\vbox{\vspace{42mm}}
\begin{center}
{\bf Symplectic tomography of Schr\"odinger cat states of a trapped ion}\\
%\vspace{9.mm}
%\hskip 6cm {{
O. V. Man'ko
%}}

{\it Lebedev Physical Institute, Moscow, Russia}
\end{center}

%\vspace{8.mm}
\begin{abstract}
The marginal distribution of squeezed, rotated and shifted quadrature for two
types of nonclassical states of a trapped ion - squeezed correlated states and
squeezed even and odd coherent states (squeezed Schr\"odinger cat states) - is
studied.


Key--words: symplectic tomography, Paul trap, Schr\"odinger cat states.

\end{abstract}

Recently, in~\cite{Vogel} it was shown that the steady state of a trapped
ion irradiated by a bichromatic laser field is a superposition of two
coherent states which is the even/odd coherent state
introduced in~\cite{Physica74} (Schr\"odinger cat
states). The theory of an ion in a Paul trap was developed
in~\cite{Glaconf,SchraMa} where the trapped ion was described by the
model of quantum oscillator with a periodically varying frequency.
In~\cite{VogRis}, a procedure was formulated to obtain the
Wigner function of the quantum system in terms of the marginal distribution of
rotated quadrature, which may be measured by a balanced homodyne detector
and the scheme called optical tomography has been used
experimentally~\cite{Raymer}. In~\cite{Mancini1}, a symplectic-tomography
procedure was suggested in which the quantum states were measured
by measuring the marginal distribution for squeezed, rotated, and shifted
quadrature. In~\cite{Mancini2}, a new equation in quantum mechanics was
introduced describing time evolution of this marginal distribution which
has completely classical form but contains all the information about
the quantum system.
The aim of this work is to consider two important types of nonclassical
states of a trapped ion (squeezed and correlated
states of the ion in a Paul trap~[3--5]
%\cite{Glaconf,SchraMa,Olya}
and
even and odd coherent states of the ion irradiated by bichromatic laser
field~\cite{Vogel}) within the framework of the symplectic-tomography
procedure and using new quantum evolution equation.

An ion in a Paul trap is described by the model of parametric oscillator.
For the trapped ion, the time--dependence of the frequency is taken to be
periodic~\cite{Glaconf}:
$$\omega ^2 (t)=1+\kappa ^2\sin ^2\Omega t\,.$$
It is easy to show that packet solutions to the Schr\"odinger equation
may be introduced and interpreted as coherent states~~$(\hbar=1)\,:$
\begin{equation}\label{ss5}
\Psi_{\alpha }(x,t)=\pi ^{-1/4}[\varepsilon (t)]^{-1/2}
\exp \left(\frac {i\dot \varepsilon (t)x^{2}}{2\varepsilon (t)}\right)
\exp \left \{
-\frac {|\alpha |^{2}}{2}-
\frac {\alpha ^{2}\varepsilon ^{*}(t)}{2\varepsilon (t)}
+\frac {{\sqrt 2}\alpha x}{\varepsilon}\right \}\,,
\end{equation}
where the classical complex trajectory satisfies the equation
$$\ddot\varepsilon(t)+\omega ^{2}(t)\varepsilon(t)=0\,,$$
with initial conditions
$$\varepsilon (0)=1\,;~~~~~\dot \varepsilon (0)=i\,,$$
where $~\alpha \,$ is a complex number.

Other normalized solutions to the Schr\"odinger equation are a squeezed even
coherent state and a squeezed odd coherent state~\cite{Physica74}
(squeezed Schr\"odinger cat states)
\begin{equation}\label{ss10} \Psi _{\alpha }^{(+)}(x,t)=2N^{(+)}\Psi
_{0}(x,t)\exp \left \{ -\frac {|\alpha |^{2}} {2}-\frac {\varepsilon
^{*}(t)\alpha ^{2}}{2\varepsilon (t)}\right \} \cosh \frac {{\sqrt 2}\alpha
x}{\varepsilon (t)}\,;
N^{(+)}=\frac {\exp (|\alpha |^{2}/2)}{2\sqrt {\cosh |\alpha |^{2}}}\,,
\end{equation}
\begin{equation}\label{ss12} \Psi _{\alpha }^{(-)}(x,t)=2N^{(-)}\Psi
_{0}(x,t)\exp \left \{ -\frac {|\alpha |^{2}}{2} -\frac {\varepsilon
^{*}(t)\alpha ^{2}}{2\varepsilon (t)}\right \} \sinh \frac {\sqrt {2}\alpha
x}{\varepsilon (t)}\,;
N^{(-)}=\frac {\exp (|\alpha |^{2}/2)}{2\sqrt {\sinh |\alpha |^{2}}}\,.
\end{equation}

In~\cite{Mancini1}, it was shown that for the  generic linear combination
of quadratures which is a measurable observable
$$\widehat X=\mu \hat q+\nu\hat p+\delta \,,$$
where $\,\hat q\,$ and $\,\hat p\,$ are the position and momentum,
respectively, the marginal distribution
$w\,(X,\,\mu ,\,\nu ,\,\delta)\,$ (normalized with respect to the $X$
variable) depending upon three extra real parameters
$\mu\,,\nu\,,\delta\,$ is related to the state of the quantum system
expressed in terms of its Wigner function $W\,(q,p)\,$ as follows
\begin{equation}\label{w}
w\,(X,\,\mu,\,\nu,\,\delta)=\int \exp \,[-ik(X-\mu q-\nu
p-\delta)]\,W(q,p)\frac{dk dq dp}{(2\pi)^2}\,.
\end{equation}
The physical meaning of the parameters $\mu,\,\nu,\,\delta \,$ is that they
describe ensemble of shifted, rotated, and scaled reference frames in
which the position $X$ is measured. This formula can be inverted and
the Wigner function of state can be expressed in terms of
the marginal distribution~\cite{Mancini1}.
In~\cite{Mancini2}, it was shown that for Hamiltonian
systems the marginal distributions satisfy the quantum
time-evolution equation, which for a trapped ion takes the form
\begin{equation}\label{TIE}
\dot w-\mu\,\frac{\partial}{\partial\nu}\,w+\omega^2(t)\nu
\,\frac{\partial}{\partial\mu}\,w=0\,.
\end{equation}
Calculating the integral~(\ref{w}) one can show that for generic
Gaussian packets of a trapped ion (also for particular case~(\ref{ss5})\,)
the marginal distribution is
\begin{equation}\label{freesolution}
w\,(X,\,\mu,\,\nu,\,\delta,\,t)=
\frac{1}{\sqrt{2\pi\sigma_X(t)}}\exp\left\{-\frac{(X-\bar {X})^2}
{2\sigma _X(t)}\right\}\,,
\end{equation}
where the dispersion of the symplectic observable $X$ and the mean
value of the observable depend on the time and the parameters as follows:
\begin{eqnarray}
\sigma _X(t)&=&{1\over2}\left(\mu^2\mid\varepsilon\mid^2+\nu
^2\mid\dot\varepsilon\mid^2 +2\mu
\nu\,\sqrt{\mid\varepsilon\dot\varepsilon\mid^2+1}\right )\,;\label{1c}\\
\overline {X}&=&\frac{\alpha}{\sqrt2}\,(\mu\varepsilon^\ast
+\nu\dot\varepsilon^\ast) +
\frac{\alpha^\ast}{\sqrt2}\,(\mu\varepsilon+\nu\dot\varepsilon)+\delta
\,.\label{2c}
\end{eqnarray}
One can check that the normalized marginal distribution~(\ref{freesolution})
with parameters~(\ref{1c}) and (\ref{2c}) satisfy the evolution
equation~(\ref{TIE}).

Now we will discuss the
marginal distribution for nonclassical states of the parametric
oscillator, namely, even and odd coherent states~\cite{Physica74}.
The Wigner function for even and odd coherent states is
\begin{eqnarray}
W_{\pm} &=& 4\mid N^{\pm}\mid^2 \exp\left\{-p^2\mid\varepsilon\mid^2-
\mid\dot\varepsilon\mid^2 q^2 +(\dot\varepsilon\varepsilon^\ast +
\varepsilon\dot\varepsilon^\ast)p q \right\}\nonumber\\
&\otimes &\left\{e^{-2\mid\alpha\mid^2}
\cosh\left(2\sqrt{2}\,[p\,\mbox{Im}\,(\alpha\varepsilon^\ast) -
q \mbox{Im}\,(\alpha\dot\varepsilon^\ast)]\right) \right. \nonumber \\
& \pm &\left. \cos\left(2\sqrt{2}\,[q\,
\mbox{Re}\,(\alpha\dot\varepsilon^\ast) - p
\mbox{Re}\,(\alpha\varepsilon^\ast)]\right)\right\}\,.\label{3c}
\end{eqnarray}
 The marginal distribution of a trapped ion in even/odd coherent
states is
\begin{equation}\label{4c}
w_{\pm}= \mid N^{\pm} \mid^2
\pi^{-1/2}\left [\mid\dot\varepsilon\mid^2\nu^2 + \mid\varepsilon\mid^2\mu^2
 +2\mu\nu\mbox{Re}\,(\dot\varepsilon\varepsilon^\ast)\right]^{-1/2} \left\{ w_1
+w_2 \pm w_3 \pm w_4\right\}\,,
\end{equation}
 where
\begin{eqnarray*}
w_1&=&\exp\left[
-\frac{\{X-\delta+2\sqrt{2}\mbox{Re}
\,(\alpha[\varepsilon^\ast\mu+\nu\dot\varepsilon^\ast])\}^2}
{\mid\dot\varepsilon\mid^2\nu^2+\mid\varepsilon\mid^2\mu^2 +2\mu\nu
\mbox{Re}\,(\dot\varepsilon\varepsilon^\ast)}\right]\,;\nonumber\\
w_2&=&\exp\left[
-\frac{\{X-\delta-2\sqrt{2}\mbox{Re}\,(\alpha[\varepsilon^\ast\mu
+\nu\dot\varepsilon^\ast])\}^2}
{\mid\dot\varepsilon\mid^2\nu^2+\mid\varepsilon\mid^2\mu^2 +2\mu\nu
\mbox{Re}\,(\dot\varepsilon\varepsilon^\ast)}\right]\,;\nonumber\\
w_3&=&\exp\left[ -2\mid\alpha\mid^2
-\frac{\{X-\delta+2i\sqrt{2}\mbox{Im}\,(\alpha[\varepsilon^\ast\mu
+\nu\dot\varepsilon^\ast])\}^2}
{\mid\dot\varepsilon\mid^2\nu^2+\mid\varepsilon\mid^2\mu^2 +2\mu\nu
\mbox{Re}\,(\dot\varepsilon\varepsilon^\ast)}\right]\,;\nonumber\\
w_4&=&\exp\left[ -2\mid\alpha\mid^2
-\frac{\{X-\delta-2i\sqrt{2}\mbox{Im}\,(\alpha[\varepsilon^\ast\mu
+\nu\dot\varepsilon^\ast])\}^2}
{\mid\dot\varepsilon\mid^2\nu^2+\mid\varepsilon\mid^2\mu^2 +2\mu\nu
\mbox{Re}\,(\dot\varepsilon\varepsilon^\ast)}\right]\,.\nonumber
\end{eqnarray*}
In the present work, we calculated the marginal distribution of
a symplectic observable (which is a generic linear quadrature) for
nonclassical states of a trapped ion modeled by a parametric
quantum oscillator. Measurements of the marginal distribution
give the possibility of measuring the quantum states. In the case of the
particular choice of the parameters
$\mu =\cos \varphi ;\, \nu =\sin \varphi,$ the measurement is reduced
to finding the marginal distribution for homodyne output and
reconstructing the Wigner function by means of the Radon transform of
the optical-tomography scheme~\cite{VogRis,Raymer}. The distribution
found for generic linear quadrature satisfies a new classical-like
equation of quantum dynamics introduced in the symplectic-tomography scheme.

{\bf Acknowledgments}

\noindent

This work was partially supported by the Russian Foundation for Basic
Research (Project No.~96--02--18623) and by the RF State Program
``Optics. Laser Physics." O.V.M. would like to express her sincere
appreciation to the Organizing Committee of the Second International
Symposium on Fundamental Problems in Quantum Physics for invitation
and Professor M. Ferrero and Dr. S.F. Huelga for hospitality.

\end{document}